\documentclass{emulateapj}
\usepackage{apjfonts}

\begin{document}


\title{Extreme Gas Fractions in Clumpy, Turbulent Disk Galaxies at $z\sim0.1$}

      

\author{David~B.~Fisher$^1$,
Karl Glazebrook$^{1,2}$,
Alberto Bolatto$^3$
Danail Obreschkow$^{2,4}$,
Erin Mentuch Cooper$^5$,
Emily Wisnioski$^6$,
Robert Bassett$^1$,
Roberto G. Abraham$^7$,
Ivana Damjanov$^8$
Andy Green$^9$
Peter McGregor$^{10}$}

\affil{$^1$Centre for Astrophysics and Supercomputing, Swinburne
  University of Technology, P.O. Box 218, Hawthorn, VIC 3122,
  Australia} \email{dfisher@swin.edu.au} \affil{$^2$ARC Centre of
  Excellence for All-sky Astrophysics (CAASTRO)} \affil{$^3$
  Laboratory of Millimeter Astronomy, University of Maryland, College
  Park, MD 29742 } 
\affil{$^4$International Centre for Radio Astronomy Research (ICRAR),
  M468, University of Western Australia, 35 Stirling Hwy, Crawley, WA
  6009, Australia} \affil{$^5$Department of Astronomy, The University
  of Texas at Austin, Austin, TX 78712, USA} \affil{$^6$Max Planck Institute for
  Extraterrestrial Physics, Garching, Germany} \affil{$^7$Department of
  Astronomy \& Astrophysics, University of Toronto, 50 St. George St.,
  Toronto, ON M5S 3H8, Canada} \affil{$^8$ Harvard-Smithsonian Center
  for Astrophysics, 60 Garden St., Cambridge, MA 02138,
  USA}\affil{$^9$Australian Astronomical Observatory, P.O. Box 970,
  North Ryde, NSW 1670, Australia} \affil{$^{10}$Research School of
  Astronomy and Astrophysics, Australian National University, Cotter
  Rd, Weston, ACT 2611, Australia}




\begin{abstract}
  In this letter we report the discovery of CO fluxes, suggesting very
  high gas fractions in three disk galaxies seen in the nearby
  Universe ($z\sim$ 0.1).  These galaxies were investigated as part of
  the DYnamics of Newly-Assembled Massive Objects (DYNAMO)
  survey. High-resolution Hubble Space Telescope imaging of these
  objects reveals the presence of large star forming clumps in the
  bodies of the galaxies, while spatially resolved spectroscopy of
  redshifted H$\alpha$ reveals the presence of high dispersion
  rotating disks. The internal dynamical state of these galaxies
  resembles that of disk systems seen at much higher redshifts
  ($1<z<3$). Using CO(1-0) observations made with the Plateau de Bure
  Interferometer, we find gas fractions of 20-30\% and depletion times
  of $t_{dep} \sim 0.5$~Gyr (assuming a Milky Way-like $\alpha_{CO}$).
  These properties are unlike those expected for low-redshift galaxies
  of comparable specific star formation rate, but they are normal for
  their high-z counterparts.  DYNAMO galaxies break the degeneracy
  between gas fraction and redshift, and we show that the depletion
  time per specific star formation rate for galaxies is closely tied
  to gas fraction, independent of redshift.  We also show that the gas
  dynamics of two of our local targets corresponds to those expected
  from unstable disks, again resembling the dynamics of high-$z$
  disks.  These results provide evidence that DYNAMO galaxies are
  local analogues to the clumpy, turbulent disks, which are often found
  at high redshift.
\end{abstract}



\section{Introduction}

In the past decade surveys have shown that there is a
decline at lower redshifts in both star formation \citep{madau1996,hopkinsbeacom2006} and
the average fraction of molecular gas mass in galaxies
\citep{tacconi2010,combes2013,carilli2013}. 
Observations show that
ionized gas in $1<z<5$ galaxies tends to be very clumpy
\cite[eg.~][]{swinbank2009,genzel2011}. These clumps are
very massive ($\sim 10^9$~M$_{\odot}$), and they are forming impressive
numbers of stars ( $\lesssim 10$~M$_{\odot}$~yr$^{-1}$). A large
fraction of clumpy galaxies have dynamics that are consistent with a
rotating disk in which the gas is also turbulent
\citep[eg.~][]{forsterschreiber2009,wisnioski2011}. 
CO observations of these galaxies indicate very high molecular gas
fractions, $f_{gas} \sim 20-50 \%$ (e.g. \citealt{tacconi2010}). Much
remains unknown about clumpy galaxies (for a recent review see
\citealt{glazebrook2013}), largely because they are almost exclusively
found at high redshift, and are thus quite difficult to observe.

Using data from the DYNAMO ({\bf DY}namics of {\bf N}ewly-{\bf
  A}ssembled {\bf M}assive {\bf O}bjects) survey, \cite{green2010}
report the discovery of a sample of galaxies at $z\sim 0.1$ whose
properties closely match those of high redshift, clumpy disk
galaxies.  The DYNAMO dataset includes integral field observations of
H$\alpha$ in 95 spiral galaxies with the highest H$\alpha$
luminosities (L$_{H\alpha}>10^{40}$~erg~s$^{-1}$) in the Sloan Digital
Sky Survey (SDSS), after excluding AGN from the sample.  Kinematic
maps of over 80\% of DYNAMO galaxies show signs of rotation, and in
over half of the sample that rotation is consistent with the
Tully-Fisher relation \citep{green2013}.  Most DYNAMO galaxies have large internal
velocity dispersions $\sigma \sim 10-100$~km~s$^{-1}$. 


If clumpy disk galaxies are dynamically unstable then a high gas
fraction is needed in these systems that also have high internal
velocity dispersions \citep[e.g.~][]{bournaud2014}.  Very high gas
fractions are observed in distant clumpy systems \citep{tacconi2013},
reinforcing the view that global Toomre instabilities may be important
in these objects.
In this letter we report the results from new
observations from Plateau de Bure Interferometer (PdBI), which confirm
high gas fractions in three DYNAMO galaxies.


\section{Methods}
\subsection{Observations \& Flux Measurements}

The DYNAMO sample was selected from the Sloan Digital Sky Survey \citep{york2000etal},
and is comprised of 95 galaxies of known stellar mass with
integral field
spectroscopy (around the H$\alpha$  line) obtained from the 
Anglo-Australian Telescope. We refer the reader to \cite{green2013} for details.

Observations centered on CO(1-0) were made for four DYNAMO
galaxies with the Plateau de
Bure Interferometer (PdBI). Three of these were recovered with significant
detections. The systems observed were D~13-5,
G~04-1, G~10-1 and H~10-2. Stellar masses for these systems
range from $1-7 \times 10^{10}$~M$_{\odot}$ and redshifts span the range
$z=0.075-0.15$. Star formation rates for D~13-5, G~10-1 and H~10-2
are $\sim$20$\pm$10 ~M$_{\odot}$~yr$^{-1}$. The star formation
rate of G~04-1 is somewhat higher at $50 \pm
10$~M$_{\odot}$~yr$^{-1}$. Galaxies G~04-1 and D~13-5 are classified
as rotating disks, while H~10-2 and G~10-1 are classified as a perturbed rotators due to
asymmetries in their velocity fields, and may be experiencing merging.

\begin{figure*} 
\begin{center}
\includegraphics[width=0.33\textwidth]{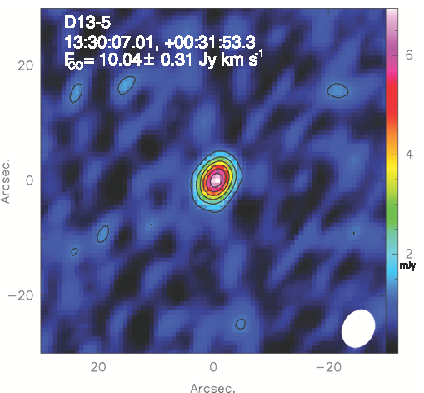}
\includegraphics[width=0.33\textwidth]{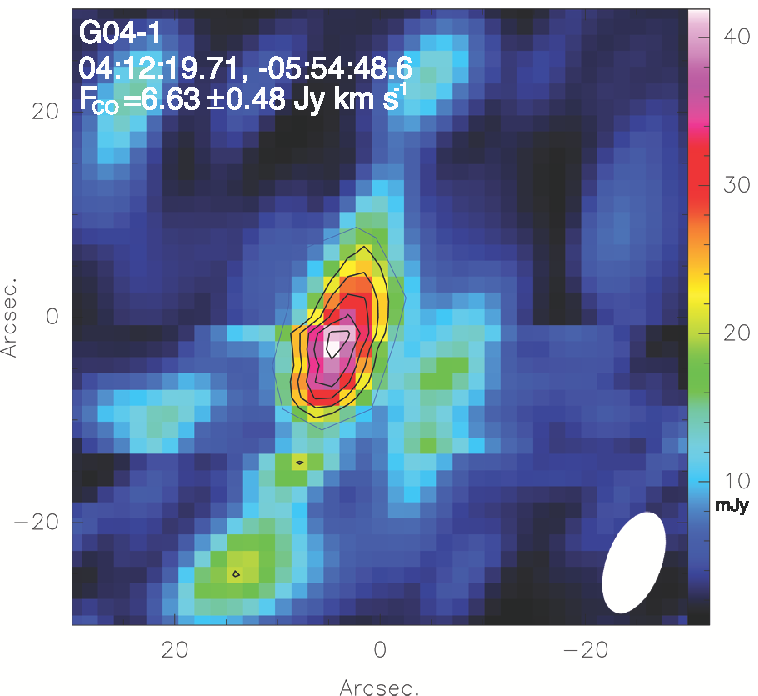}
\includegraphics[width=0.33\textwidth]{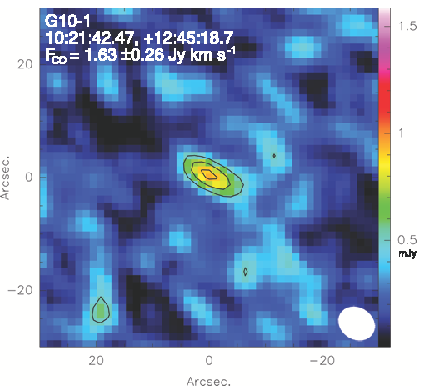} \\
\includegraphics[width=0.33\textwidth]{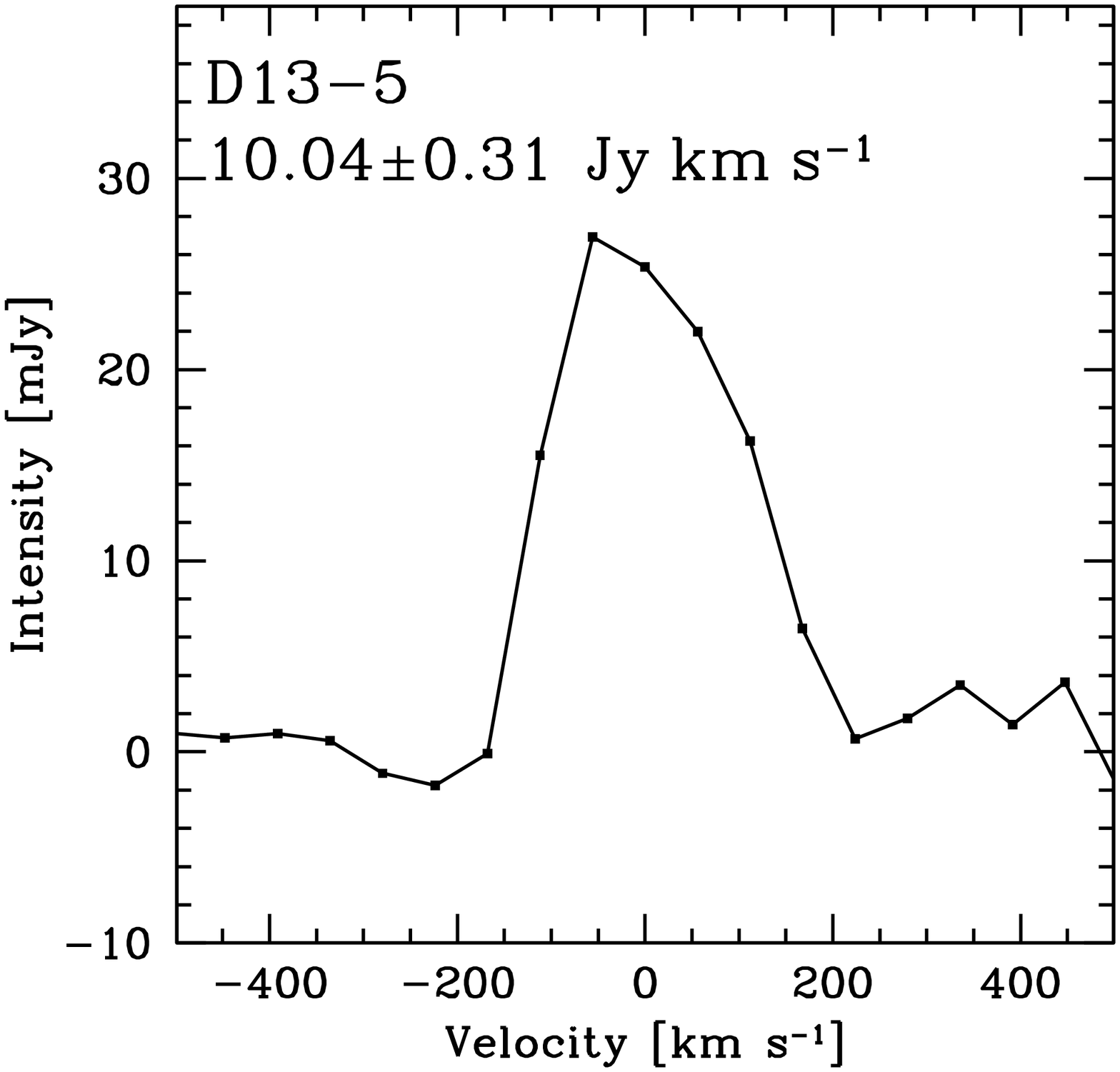}
\includegraphics[width=0.33\textwidth]{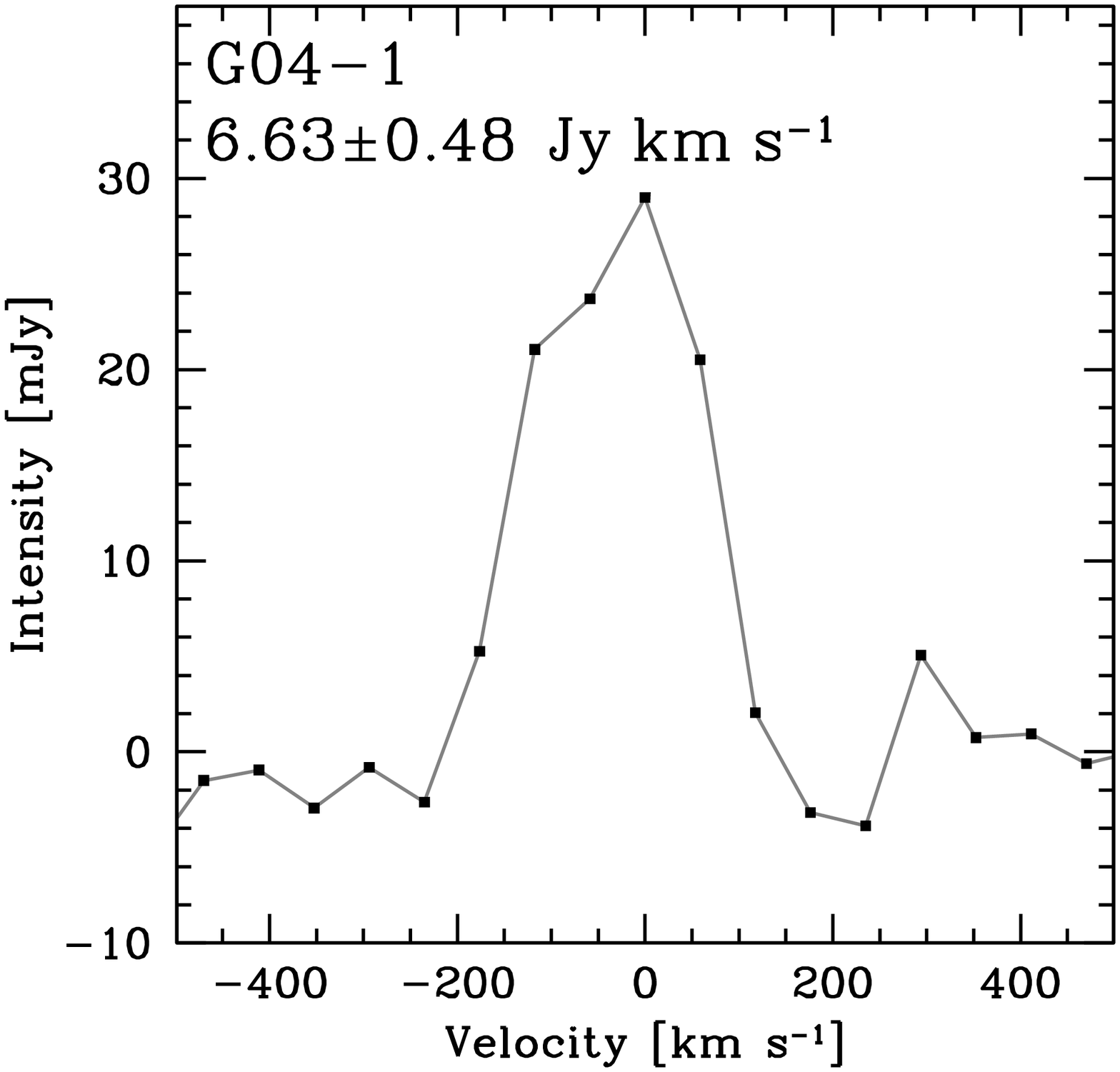}
\includegraphics[width=0.33\textwidth]{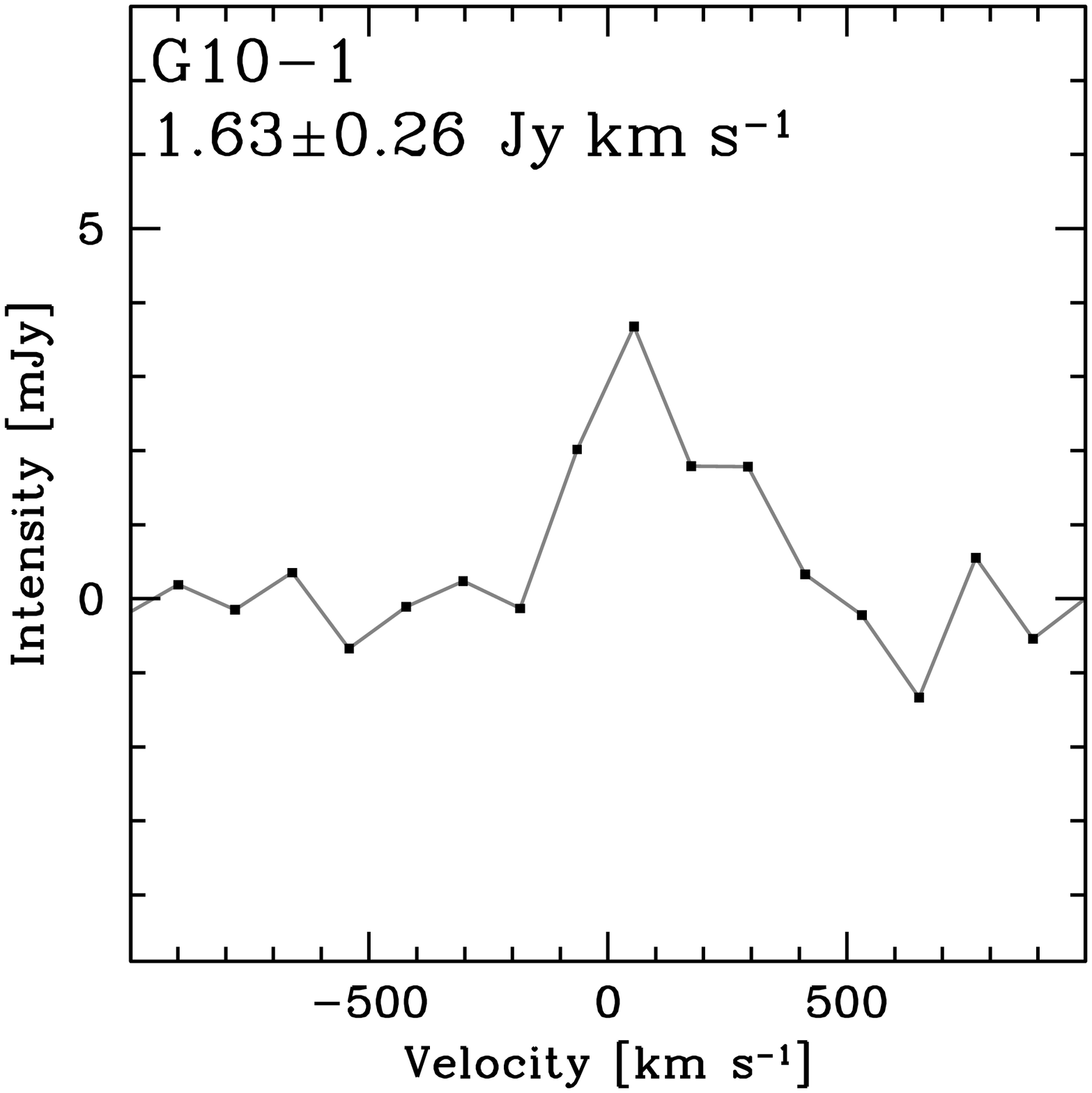}
\end{center}
\caption{ Top row: Moment 0 maps of the three galaxies in which we
  detect CO(1-0) emission. (We do not recover a statistically
  significant flux on H~10-2).  The typical beam size of our
  observations is 6 arcseconds, beam shapes are shown as the white
  ellipse in the bottom right corner of each panel. The beam of G~04-1
  is very elongated due to the galaxy being lower on the sky.  Note
  that in all cases the CO emission is unresolved, the elongated
  structure of G~04-1 is due to the low elevation of the
  observation. The countours are in units of 2$\sigma$. Bottom row:
  Spectra of each detection.  In these 3 galaxies, the maps and spectra show 
  clear detections in CO(1-0).  \label{fig:maps}}
\end{figure*}

Observations described here were made using the PdBI in 
D configuration (FWHM $\sim 6$'') from 30-May-2013
to 16-July-2013. On-source integration times were 1-2 hours per
target.  Data were calibrated and reduced with standard methods at
IRAM. 
The spectra were binned into 20~km~s$^{-1}$
channels. Fluxes were measured, using standard GILDAS routines, by
defining an ellipse in the moment zero CO(1-0) map (the outer contour
in Fig.~\ref{fig:maps}). The noise levels are 0.31~Jy~km~s$^{-1}$
(D~13-5), 0.48~Jy~km~s$^{-1}$ (G~04-1), 0.26~Jy~km~s$^{-1}$ (G~10-1)
and 0.42~Jy~km~s$^{-1}$ (H~10-2). 

\begin{figure} 
\begin{center}
\includegraphics[width=0.5\textwidth]{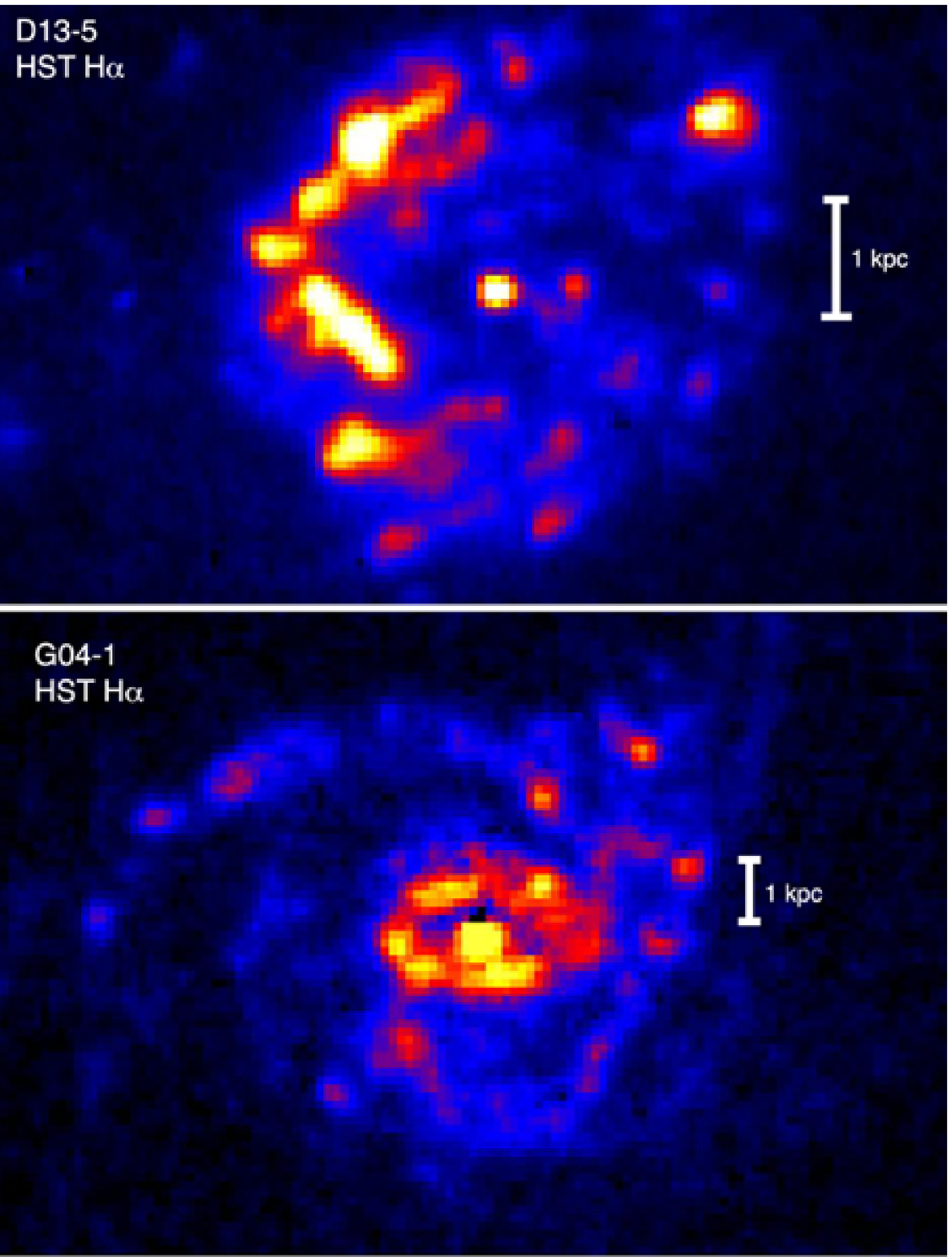}
\end{center}
\caption{HST H$\alpha$ maps of DYNAMO galaxies reveal the presence of
  massive star forming clumps. Here we show two HST H$\alpha$ maps for
  our target disk galaxies D~13-5 (top) and G~04-1 (bottom). In each
  figure the white line indicate 1~kpc. The clumps are of high~S/N and
  have comparable SFR to those observed at high redshift.       \label{fig:clumps}}
\end{figure} 

On three of our sources we recovered significant detections. 
We do not find a statistically
significant flux for H~10-2, our most distant source, 
but find an upper limit consistent with the gas
fractions measured in the other three galaxies (see Section 3). 
Data for each detected galaxy is shown Fig.~\ref{fig:maps}. The flux
measurements on D~13-5, G~04-1, G10-1 have $S/N\sim $30, 13 and 6,
respectively.

In Fig.~\ref{fig:clumps} we show H$\alpha$ maps of two of our target
galaxies, D~13-5 and G~04-1, obtained using the Advanced Camera for
Surveys (ACS) ramp filters on the Hubble Space Telescope (HST; PID
12977, PI Damjanov). The DYNAMO galaxies shown have a clumpy
distribution of ionized gas, and their appearance in H$\alpha$ is
remarkably similar to that seen in many high-redshift galaxies
\citep[e.g.][]{genzel2011,wisnioski2012} and simulations of clumpy
galaxies \citep{bournaud2014}. These galaxies both show evidence of a
ring of gas, similar to that of high-$z$ galaxies
\citep{genzel2008,genzel2011,genzel2014}. Though we note that gas rings are
common in low-$z$ non-turbulent, often barred, galaxies as well
\citep[eg.][]{boker2008}.  Individual clumps have very bright
H$\alpha$ emission, implying star formation rates of individual clumps
$\sim 1-10$~M$_{\odot}$~yr$^{-1}$. Surveys of high-$z$ lensed
galaxies, in which the clump sizes are not limited by resolution, find
that clumps are typically 100-400~pc \citep{jones2010}. This is
consistent with the sizes of clumps in our target galaxies. A
detailed analysis of the clump properties from a larger sample of HST
images will be the subject of a future paper (Fisher et al. {\em in
  prep}).

\subsection{Conversion to total molecular gas }

The conversion of CO(1-0) flux to total molecular gas mass was done in
the usual fashion, where 
\begin{equation}
M_{mol} = \alpha_{CO}\, L_{CO}.
\end{equation}
The quantity $L_{CO}$ represents the luminosity of CO(1-0), and $\alpha_{CO}$
is the conversion factor of CO-to-H$_2$ (including the 1.36$\times$
factor accounting for Helium). Typical values for
$\alpha_{CO}$ range from
0.8-4.5~M$_{\odot}$~pc$^{-2}$~(K~km~s$^{-1}$)$^{-1}$. 
For discussion of this conversion factor in different
environments see
\citealp{youngscoville1991,downes1998,fisher2012submitted}
and \cite{bolatto2013} for a detailed review.

\begin{figure*}[t]
\begin{center} 
\includegraphics[width=0.99\textwidth]{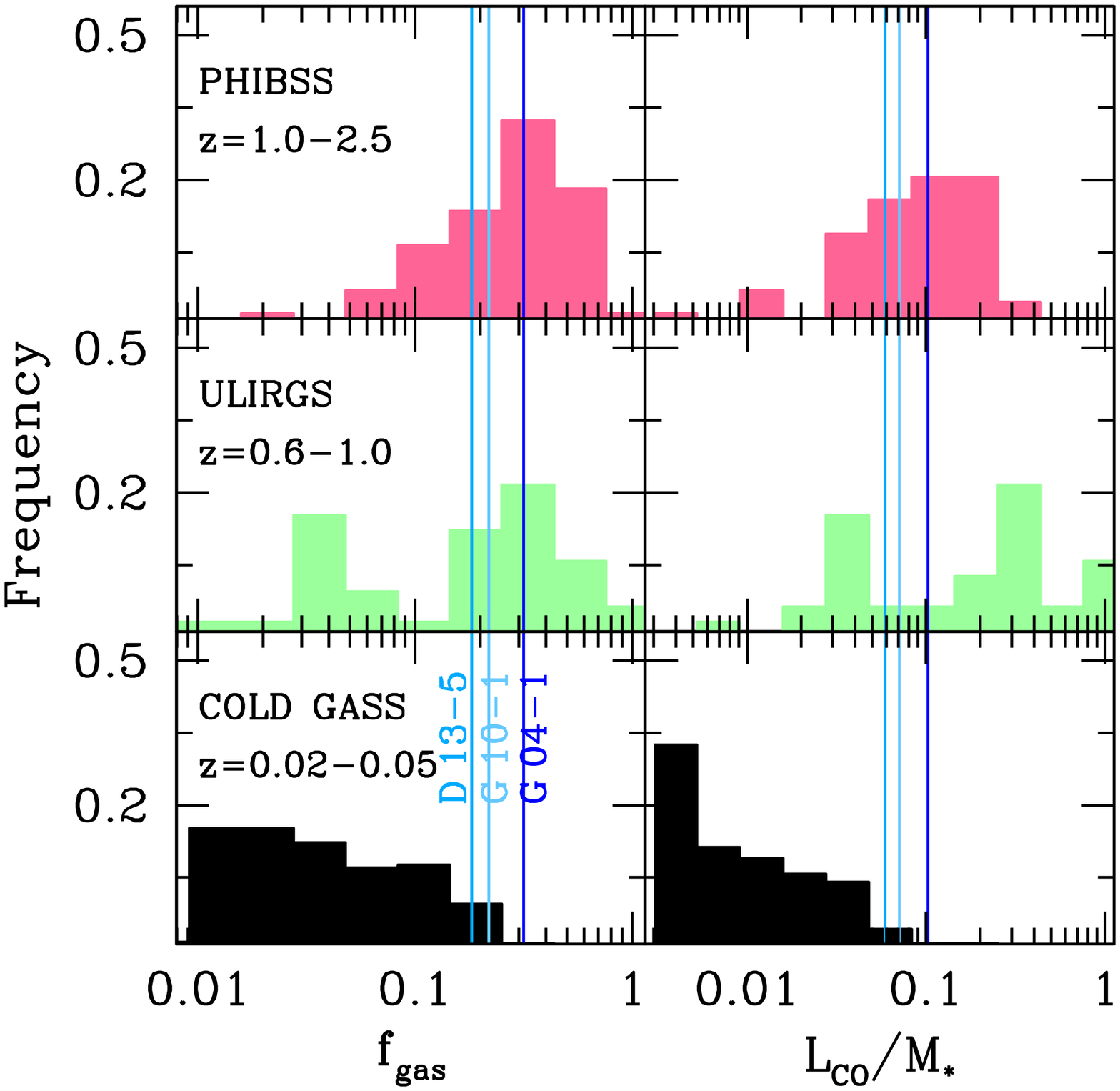}
\end{center}
\caption{A comparison of the gas fractions ($f_{gas} =
  M_{mol}/(M_{mol}+M_{star})$) of our target galaxies to the
  distribution of gas fractions from three surveys at different
  redshifts. In the right panels we show a similar quantity the ratio
  of CO luminosity to stellar mass for the same galaxies, a quanity
  that is independent of assumptions on $\alpha_{CO}$. From bottom to
  top, histograms correspond to the GASS survey \citep{saintonge2011a}
  of low redshift galaxies, $z=0.2-1.0$ URLIRGs from \cite{combes2013}
  , and the $z=1.0-2.5$ PHIBSS survey \citep{tacconi2013}. The gas
  fractions of our DYNAMO galaxies are indicated by three vertical
  lines. From left to right, these correspond to D~13-5 (cyan), G~10-1
  (light blue) and G~04-1 (dark blue). The DYNAMO galaxies, especially
  G~04-1, have high gas fractions compared to other galaxies at
  $z=0-0.1$, and are much more consistent with star forming galaxies
  that are found at high redshifts. \label{fig:hist} }

\end{figure*}

G~04-1 and D~13-5 have H$\alpha$ kinematics that indicate rotating
systems \citep{green2013}, and the surface brightness profile of the
star light in SDSS images closely resembles an exponential disk with
at most a small (B/T<10\%) bulge.  It therefore seems reasonable to
assume a conversion factor similar to the Milky Way \cite[as done
by][]{tacconi2013}. However, the large star formation rates in these
galaxies suggest that a smaller conversion factor ($\alpha_{CO} \sim
1$) may be appropriate. \cite{bolatto2013} provides a first estimate
for a universal equation to determine the CO-to-H$_2$ conversion
factor in galaxies, such that
\begin{equation}
\alpha_{CO} \approx 2.9 \exp\left ( \frac{0.4}{Z' \Sigma^{100}_{GMC}}
\right ) \left ( \frac{\Sigma_{total}}{100 M_{\odot}~pc^{-2} }\right
)^{-0.5}
\label{alphaeqn}
\end{equation}

\noindent where $Z'$ represents the metallicity in solar units,
$\Sigma^{100}_{GMC}$ represents the average surface density of
molecular clouds in units of 100~M$_{\odot}$~pc$^{-2}$, and
$\Sigma_{total}$ is the total surface density of the region in
question. 
We find that metallicities fall in the range $Z'\sim 0.9-1.1$,
using the [NII]/H$\alpha$ ratio \citep{pettini2004} from SDSS
spectra. The average surface density of molecular
clouds in these clumpy galaxies is presently uncertain. Star forming
clumps may be as massive as $10^9$~M$_{\odot}$ and have radii
$R_{clump}\sim 500$~pc \citep{swinbank2012}, which yields
$\Sigma^{100}_{GMC}\sim 10$. The total surface densities of our
galaxies are 100-300~M$_{\odot}$~pc$^{-1}$, measured from SDSS surface
photometry. Inserting these quantities into Equation~\ref{alphaeqn}
results in $\alpha_{CO} \sim 3.1
$~M$_{\odot}$~pc$^{-2}$~(K~km~s$^{-1}$)$^{-1}$ for
our targets. This value is adopted for $\alpha_{CO}$. We note that our
chosen $\alpha_{CO}$ is consistent with estimates of $\alpha_{CO}$ in
the Milky Way \citep[eg.~][]{dame2001}.

We do not currently have sufficient data to measure the temperature of
the gas in these galaxies, though indirect evidence supports our
choice of disk-like $\alpha_{CO}$. Mentuch-Cooper et al. ({\em in
  prep}) find with $Herschel$ data that in similar DYNAMO galaxies the
dust temperatures are $T_{dust}\sim25-35$~K, which is consistent with
$\alpha_{CO} \sim3-4$ \citep{magnelli2012xco}. Also, connections between ISM
state and mid-plane pressure in galaxies \citep{blitzrosolowsky2006},
suggest a link between gas temperature and total surface density. It
is therefore likely that using surface density implicitly accounts
for gas temperature. Nonetheless, in light of the well-known
uncertainties of CO-to-H$_2$ conversion in such high star formation
systems, for our main result (Fig.~\ref{fig:hist}), we show
$L_{CO}/M_{star}$ along with the more physical gas fraction.

We use variable $\alpha_{CO}$ from \cite{bolatto2013}
to adjust the gas masses of our comparison samples. For the COLD GASS
galaxies we estimate the stellar surface density with stellar masses
and sizes given in \cite{saintonge2011a}; we assume
$\Sigma^{100}_{GMC}\approx 1.7$ \citep{bolatto2008} and $Z'\approx 1$.
For \citep{saintonge2011a} galaxies the typical adjustment is very
minor. The ultra luminous infrared galaxies (ULIRGs) from
\cite{combes2013} require $\alpha_{CO}=0.8$, and thus these gas masses
do not change.  

We make the same assumptions about $\Sigma^{100}_{GMC}$ in PHIBSS
galaxies as for DYNAMO galaxies, and use published values
\citep{tacconi2013} for stellar mass and size to estimate the stellar
surface density.  Gas masses in gas-rich star forming galaxies, are not known to better than a factor
of $\sim 2\times$ due to uncertainties in the conversion factor.  For
the PHIBSS galaxies the gas masses are reduced by 10-40\%, by using
the variable conversion factor. 

\begin{figure*}
\begin{center}
\includegraphics[width=0.89\textwidth]{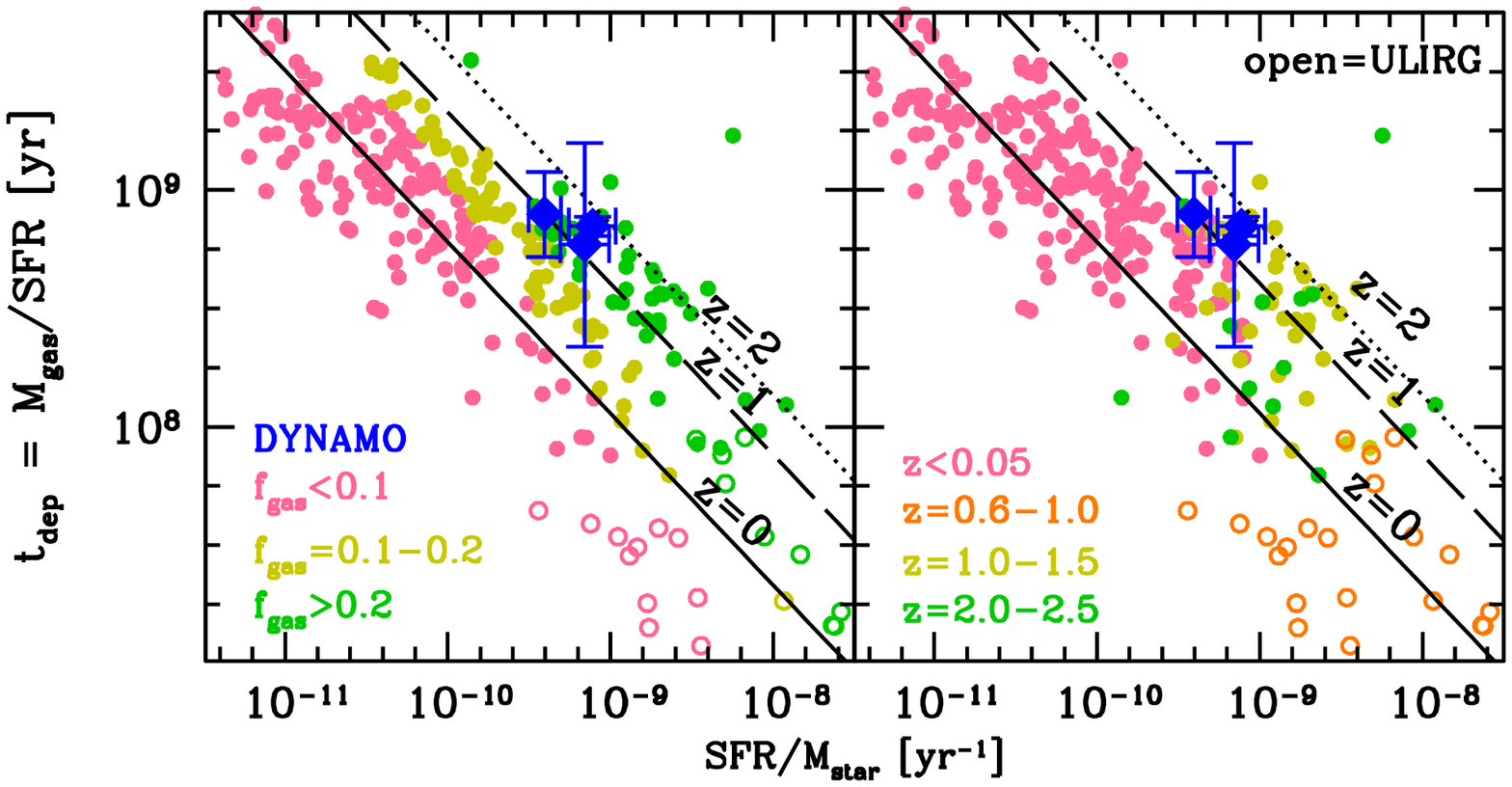}
\end{center}
\caption{ The star formation in DYNAMO galaxies is consistent with
  that of other gas rich main-sequence galaxies. This figure compares
  the depletion time (M$_{gas}$/SFR) to the specific star formation
  rate. Open symbols represent ULIRGs, while closed symbols are main-sequence disks.  In
  the left panel we organize the symbols based on molecular gas
  fraction. In the right panel symbol colors are based on
  redshift. The solid line is the relationship from the COLD GASS
  survey \citep{saintonge2011a}, and the higher redshift lines simply
  adjust that relation by a constant offset based on the measured
  evolution of the specific SFR with redshift
  \citep{perez2008}.
  \label{fig:tdep}}
\end{figure*}

\section{Results}

The gas mass fraction $f_{gas}$, defined as
$f_{gas} = M_{mol}/(M_{mol}+M_{star})$, of our
sample is
presented in Fig.~\ref{fig:hist}. The
DYNAMO galaxies have $f_{gas}=0.18\pm0.07$,
$0.31\pm 0.03$ and $0.22\pm0.21$ for D~13-5, G~04-1, and G~10-1
respectively. An upper-limit to the gas fraction of H~10-2 is
$f_{gas}<0.22$. As it adds essentially no information, the upper-limit
of H~10-2 is not displayed in our figures.

Alternatively, if one assumes a starburst like $\alpha_{CO}\sim 1$ the
gas fractions are $\sim$0.06, 0.10, and 0.07 for D~13-5, G~04-1, and G~10-1.


From the bottom panel of Fig.~\ref{fig:hist} it can be seen that our
DYNAMO galaxies
are more gas rich than typical low redshift galaxies.  The
histogram in this panel corresponds to the $z\sim
0$ COLD GASS survey \citep{saintonge2011a}, which is
volume limited
for galaxies with
$M_{star}>10^{10}$~M$_{\odot}$. \cite{saintonge2011a} finds that only
5\% of low-$z$ galaxies have gas fractions greater than 15\% and the
median gas fraction for COLD GASS galaxies in which CO is detected is
7\% with a standard deviation of $\sim 6$\%.


In the middle panel of Fig.~\ref{fig:hist} the gas fractions
of our target galaxies are compared with those of the ULIRG
sample presented by \cite{combes2013}, who measured the gas
fractions and star formation efficiency for systems with
$L_{FIR}>10^{12}$~L$_{\odot}$, spanning the redshift range
$z=0.6-1.0$. Note that \cite{combes2013} published stellar masses
using the Salpeter IMF, so we multiplied these stellar masses by
0.55 to convert them to the \cite{chabrier2003} IMF, thus making
them consistent with other stellar masses used in this paper. The
ULIRG sample has a large range of gas fractions, and in
Fig.~\ref{fig:hist} shows a double-peaked distribution. The gas
fractions of our DYNAMO galaxies fall close to the middle of the range
spanned by the more gas rich part of the ULIRG sample.

In the top panel of Fig.~\ref{fig:hist}, we show that the gas
fractions of our target DYNAMO galaxies are quite similar to
those of the
main-sequence $z=1-3$ star forming galaxies 
presented by \cite{tacconi2013} 
as part of the
PHIBSS survey. PHIBSS  targets star forming and
starbursting galaxies of the high redshift Universe, many
of which have been revealed to clumpy,
turbulent systems. 

In the right panels of Fig.~\ref{fig:hist} we show that the ratio of
CO luminosity-to-stellar mass, $L_{CO}/M_{*}$. The results we find here
are similar to the right panels, in which we estimate the gas mass. We
find that DYNAMO galaxies are at the extreme high end of
$L_{CO}/M_{*}$ values when compared to the GASS survey. Also, we find
that luminosity of CO gas per unit stellar mass from DYNAMO galaxies is
consistent with the center of the distribution of that of PHIBSS
galaxies. The high gas fractions, and CO flux ratios, in DYNAMO
galaxies are not due to low stellar masses, which span the range from
$3-7\times 10^{10}$~M$_{\odot}$, similar to stellar masses found in
gas rich PHIBSS galaxies, and much higher than local gas rich dwarfs.

%

The depletion times ($t_{dep}=M_{gas}/SFR$) of the DYNAMO galaxies
observed are $0.4 - 0.6$ Gyr, assuming $\alpha_{CO}$ from
Equation~\ref{alphaeqn}. ULIRGS are known to have short depletion
times \citep[$\sim 0.01$~Gyr,][]{carilli2013} when compared to that of
``main-sequence'' mode disks \citep[$\sim 1$~Gyr,
eg.][]{leroy2008,rahman2012}. Note, that if we assume a lower
CO-to-H$_2$ conversion factor $\alpha_{CO}=1$~(K~km~pc$^{-1}$)$^{-1}$,
the depletion times are $0.1 - 0.2$~Gyr.  This value is outside the
range spanned by the \cite{combes2013} ULRIG comparison sample (albeit
within error bars), and an order of magnitude longer than the median
depletion time in the ULIRG sample.



The relationship between $t_{dep}$ and the specific star-formation
rate of a main-sequence disk galaxy changes with redshift
\citep{saintonge2011b}. 
In the left panel of Fig.~\ref{fig:tdep} we show that there is a
strong tendency for galaxies with high gas fractions to have a longer
depletion time for a given $SFR/M_{star}$, including our three
galaxies with $f_{gas}\sim 20-30$\%. This is not surprising
\citep[see][]{tacconi2010}, as $t_{dep} \times SFR/M_{star} =
M_{gas}/M_{star}$. The correlation with redshift is weaker by
comparison, and in the right hand panel we show that the nearby DYNAMO
galaxies are clear outliers to the trend, falling in the range
populated by $z=1-2$ galaxies. Also, $\sim20$\% (11/51) of galaxies
with $z>1$ are located within the spread of the $t_{dep}$-specific SFR
relationship for $z=0$ (COLD GASS) galaxies.

In summary, we find that the nearby $z\sim 0.1$ DYNAMO galaxies we
target are very similar to the star forming galaxies observed by
\cite{tacconi2013}, which are found at $z\sim1.0-2.5$, in both their gas
fractions and depletion times.

\section{Discussion} 


In this paper we report observations that suggest high gas fractions in three
nearby clumpy, turbulent galaxies from the DYNAMO sample. Assuming
a disk-like CO-to-H$_2$ conversion,  the
molecular gas fractions ($f_{gas} = M_{mol}/(M_{mol}+M_{star})$) of
these three galaxies are $f_{gas}=20-30$\%. Compared to other
low-redshift galaxies, the DYNAMO galaxies are strong outliers: they
are gas-rich like $1<z<2.5$ galaxies, but they are located in the
local Universe. As we show in Fig.~\ref{fig:tdep}, the displacement of
galaxies, including our target galaxies, from the
$t_{dep}-SFR/M_{\star}$ relations defined by the local COLD GASS
survey is more strongly correlated with gas fraction than it is to
redshift. We remind the reader of the caveat that this result depends
on the assumption  of the disk-like $\alpha_{CO}$ for DYNAMO and PHIBSS
galaxies.  A straightforward interpretation of our results is that,
independent of redshift, having very large gas fraction is a crucial
factor to determining a galaxies star forming properties, such as the
presence of massive star forming clumps.

Giant star forming clumps, like those in Fig.~\ref{fig:clumps}, may be
associated with local gravitational instabilities
\citep{genzel2011,glazebrook2013} in disks that are marginally stable
\citep[see also][]{dekel2009,bournaud2009}. 
Under the assumption that G~04-1 and D13-5 are disks, and that their
clumps are due to turbulent gravitational instabilities, we expect
that these galaxies would be unstable as well, and thus have $Q<1$.
We note that an alternate hypothesis, in which G~04-1 and D~13-5 are
the result of merging, this calculation would have less meaning, and
$Q$ values would be affected by the change to the gas fractions,
discussed in \S~2~\&~3.  Similar to \cite{genzel2011}, we express the
\cite{toomre1964} $Q$ parameter as a function of the gas fraction,
rotation velocity (V), and velocity dispersion ($\sigma$) as
follows\footnote{For more discussion regarding the measurement of $Q$
  in disks, see \cite{leroy2008}, \cite{vanderkruit2011} and
  \cite{glazebrook2013}.}:
\begin{equation}
Q=a \left ( \frac{\sigma}{V} \right ) \left ( \frac{1}{f_{gas}} \right
) \label{eq:q}.
\end{equation}
In the local Universe, typical disks are stable, with $Q\gtrsim 2$
\citep{vanderkruit2011}. Taking the dynamical quantities from
\cite{green2013}, assuming $a=1$ for a Keplerian disk and the
non-merger conversion CO-to-H$_2$ factor, we measure $Q\sim 0.8$ for
D~13-5 and $Q\sim 0.6$ for G~04-1. Assuming these galaxies are truly
disks, these results are consistent with a picture in which the clumps
observed in DYNAMO galaxies are likely the result of unstable gas.

We conclude by noting the continued similarity between DYNAMO galaxies
and turbulent disks seen at high redshifts.  Many DYNAMO galaxies have
internal dynamical structures similar to those observed in $z\sim 1-2$
galaxies \citep{forsterschreiber2009}: they are rotating disks with
high internal velocity dispersions \citep{green2010,green2013}. We now
add that observations indicate clumpy H$\alpha$ morphologies, high gas
fractions, and that they lie on the same portion of the depletion time
vs. specific star formation rate diagram occupied by star-forming
galaxies at high redshift.  Taken together, the evidence continues to
be consistent with the hypothesis that DYNAMO galaxies are indeed
close analogues to the clumpy galaxies seen at high redshift.  In
future papers our group will exploit the proximity of DYNAMO to study
both high spatial resolution properties of galaxies with massive star
forming clumps (Fisher et al.  {\em in prep.}) and the faint stellar
kinematics seen in these same galaxies (Bassett et al. {\em
  submitted}).
 
\acknowledgments

We thank the referee for their helpful comments. We are very grateful to
Sabine K\"onig for observational support and help calibrating the
data. This work is based on observations carried out with the IRAM
Plateau de Bure Interferometer. IRAM is supported by INSU/CNRS
(France), MPG (Germany) and IGN (Spain).

DBF acknolwedges support from Australian Research Council (ARC)
Discovery Program (DP) grant DP130101460.  Support for this project is
provided in part by the Victorian Department of State Developement,
Business and Innovation through the Victorian International Research
Scholarship (VIRS).  AB acknowledges partial support form AST
08-38178, CAREER award AST 09-55836, and a Cottrell Scholar award from
the Research Corporation for Science Advancement.


\end{document}